# *i*-Caloric Effects: a proposal for normalization

**William IMAMURA**[(a,b)], **Lucas S. PAIXÃO**[(a)], **Érik O. USUDA**[(a,c)], **Nicolau M. BOM**[(a)], **Sergio GAMA**[(c)], **Éder S. N. LOPES**[(b)], **Alexandre M. G. CARVALHO**[(a,d)]

[(a)] Laboratório Nacional de Luz Síncrotron (LNLS), Centro Nacional de Pesquisa em Energia e Materiais (CNPEM), Campinas-SP, CEP 13083-100, Brazil
[(b)] Faculdade de Engenharia Mecânica, UNICAMP, Campinas-SP, CEP 13083-860, Brazil
[(c)] Universidade Federal de São Paulo, UNIFESP, Diadema-SP, CEP 09913-030, Brazil
[(d)] Departamento de Engenharia Mecânica, Maringá-PR, CEP 87020-900, Brazil
To whom correspondence may be addressed: WI (williamimamura@yahoo.com.br)

**ABSTRACT**

Solid-state cooling based on *i*-caloric effects is considered the most promising alternative to replace the conventional vapor-compression refrigeration systems. It is possible to define an *i*-caloric effect as a thermal response registered in a material upon the application of an external field, characterized by an adiabatic temperature change ($\Delta T_S$) or an isothermal entropy change ($\Delta S_T$). Depending on the nature of the external field (magnetic field, electric field or stress field), the *i*-caloric effects can be categorized as magnetocaloric effect, electrocaloric effect, and mechanocaloric effect. We can still subdivide mechanocaloric effect in: elastocaloric effect, driven by uniaxial stress; barocaloric effect, driven by isotropic stress (pressure); and torsiocaloric effect, driven by a torque in a prismatic bar, causing a pure shear stress of torsion. The study of *i*-caloric effects dates from the beginning of 19$^{th}$ century. Nevertheless, due to the independent development of investigations on each effect, there are no stablished standards regarding terminology or results evaluation up to now, making the understanding quite challenging for the community. In this context, we present a proposal for normalization of *i*-caloric effects, considering different aspects, such as nomenclature, thermodynamics and figures of merit.

Keywords: *i*-caloric effects, Normalization.

## 1. INTRODUCTION

Modern refrigeration devices rely on vapor-compression cycles, a technology that dates to the early development of thermodynamics and can be thought as a barocaloric effect. Due to energy efficiency and environmental issues, alternative technologies are under development, and one promising option is solid-state cooling based on *i*-caloric effects.

The general definition of *i*-caloric effect can be stated as a thermal response when a material is exposed to a change in external fields (where "*i*" stands for intensive thermodynamic variable – denoting the external fields). The nature of the response depends on the thermodynamic process performed on the material. The effects are characterized by: the temperature change ($\Delta T_S$), when the material undergoes an adiabatic process; or the entropy change ($\Delta S_T$), when the material undergoes an isothermal process.

*i*-caloric effects have a somewhat erratic timeline, since research on each *i*-caloric effect was conducted independently. The first solid-state *i*-caloric effect observed was the elastocaloric effect in 1805, when J. Gough (1805) reported the temperature change of natural rubber under rapid stretching. A few decades later, W. Thomson (The Lord Kelvin) used thermodynamic considerations to predict the mechanocaloric effect (Thomson, 1855), and later the magnetocaloric (Nichol, 1860), and the electrocaloric effects (Thomson, 1878). Despite the theoretical predictions, experimental observations of each *i*-caloric effect occurred years apart from each other. In 1917, P. Weiss and A. Piccard (1917) reported the first observation of the magnetocaloric effect;

only then the current terminology began to appear, as the authors coined the term magnetocaloric (from French, "*magnétocalorique*"). About a decade later, P. Debye (1926) and W. Giauque (1927) were, independently, investigating pathways to reach low temperatures using the adiabatic demagnetization of paramagnetic salts. The experimental realization of such process was reported by W. Giauque and D. MacDougall (1933). Meanwhile, the electrocaloric effect was first observed on Rochelle salt, and reported by P. Kobeko and J. Kurtschatov (1930).

After years of steady development, magnetocaloric effect gained new visibility in 1976, when G. Brown (1976) reported the first prototype of a magnetic refrigerator working near room temperature. A great deal of interest followed the reports of giant *i*-caloric effects. First the giant magnetocaloric effect in $Gd_5Si_2Ge_2$ compound in 1997 (Pecharsky and Gschneidner, Jr., 1997), followed by the giant electrocaloric effect in thin film $PbZr_{0.95}Ti_{0.05}O_3$ in 2006 (Mischenko et al., 2006), and the giant barocaloric effect in the Ni-Mn-In shape-memory alloy in 2010 (Mañosa et al., 2010). Recently, multiferroic materials merged effects independently studied until recently, with the promise of even larger temperature changes.

Numerous findings were not mentioned on the short account presented above. Also, there are reviews that cover in detail the history of the *i*-caloric effects (Lu and Liu, 2015; Tishin et al., 2016; Valant, 2012). The current proposal is partly intended to avoid ambiguity in the common terminology among multiple effects. This point is especially important since the rise of multiferroics as potential *i*-caloric materials. Another goal is to discuss best practices when comparing results for different materials, obtained from different protocols.

## 2. *i*-CALORIC EFFECTS

### 2.1. Types of *i*-caloric effects

Depending on the nature of the external field (magnetic field ($H$), electric field ($E$), or stress field ($\sigma$)), it is possible to categorize the *i*-caloric effects as magnetocaloric (*h*-CE), electrocaloric (*e*-CE), and mechanocaloric ($\sigma$-CE), as illustrated in Fig. 1 and organized in Table 1.

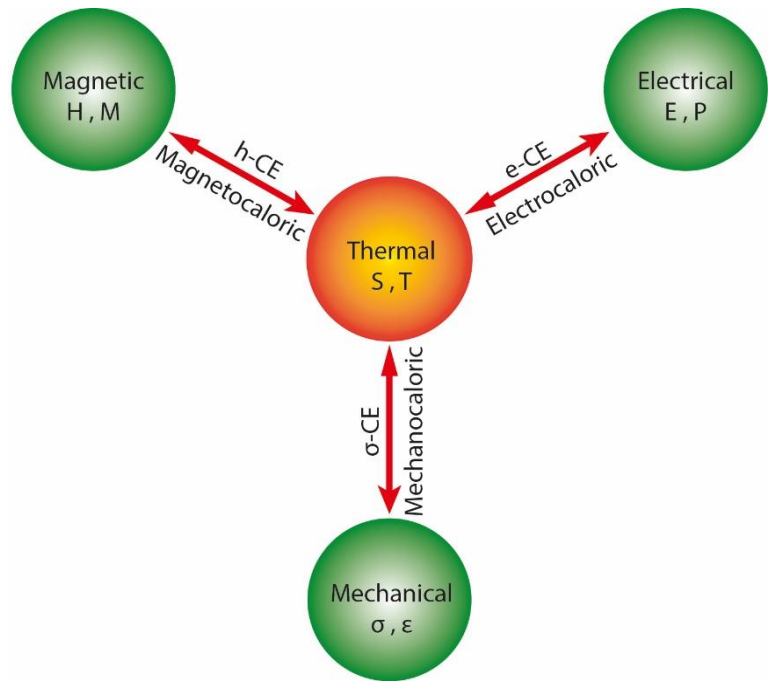

**Figure 1: Scheme for *i*-caloric effects. The red arrows indicate the possible thermal and entropic responses. Adapted from Vopson (2013).**

**Table 1: External field ($X$) and its conjugated intensive variable ($Y$) for each type of system and the corresponding $i$-caloric effect.**

| System | $X$ | $Y$ | $i$-caloric effect |
|---|---|---|---|
| Magnetic | Magnetic field ($H$) | Magnetization ($M$) | Magnetocaloric ($h$-CE) |
| Electrical | Electric field ($E$) | Polarization ($P$) | Electrocaloric ($e$-CE) |
| Mechanical | Stress field ($\sigma$) | Strain* ($\varepsilon$) | Mechanocaloric ($\sigma$-CE) |
|  | Uniaxial stress |  | Elastocaloric ($\sigma_e$-CE) |
|  | Isostatic stress |  | Barocaloric ($\sigma_b$-CE) |
|  | Pure shear stress |  | Torsiocaloric ($\sigma_t$-CE) |

* If $Y$ is an intensive variable in units of mass, the strain must be divided by a reference density ($\rho_0$).

Here, it is important to emphasize and clarify that the $\sigma$-CE is a generic terminology for $i$-caloric effects driven by a stress field change. Particular cases are classified according to the non-zero components of the Cauchy stress tensor ($\sigma = \sigma_{ij}$). For example, the elastocaloric effect ($\sigma_e$-CE; Fig. 2a) is a manifestation of a thermal response driven by a uniaxial stress ($\sigma_{ij} = 0$ for $i \neq j$, and there is only one $\sigma_{ij} \neq 0$ for $i = j$). The barocaloric effect ($\sigma_b$-CE; Fig. 2b) is driven by the principal stresses equal to pressure ($p$) with no shear stresses ($\sigma_{xx} = \sigma_{yy} = \sigma_{zz} = p$, and $\sigma_{ij} = 0$ for $i \neq j$). Finally, the torsiocaloric effect ($\sigma_t$-CE; Fig. 2c) is driven by a pure shear stress in a prismatic bar; for a torsion about an arbitrary $k$-axis, the shear-stress applied on the cross-section perpendicular to the $k$-axis is non-zero, as well as their symmetrical components ($\sigma_{ij} = 0$ for $i = j$, and $\sigma_{ij} = 0$ for $i$ and $j \neq k$).

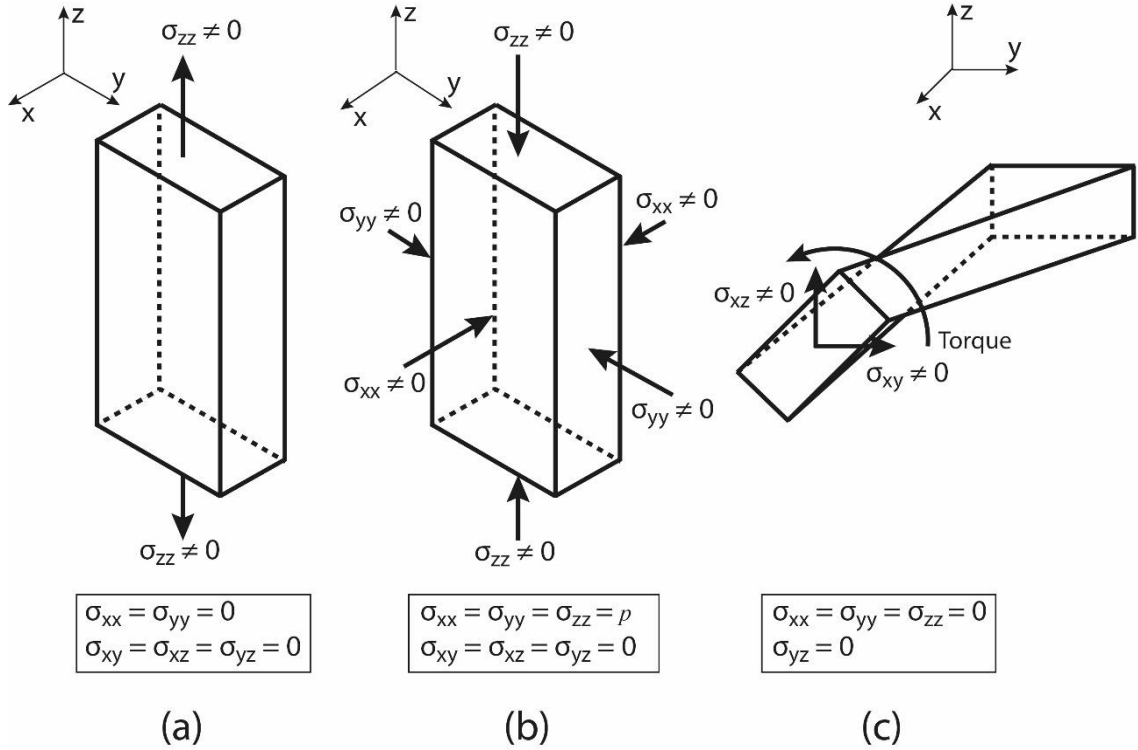

Figure 2: Mechanocaloric effects: (a) elastocaloric effect; (b) barocaloric effect; and (c) torsiocaloric effect.

## 2.2. Synonyms and other terminologies

Multiple terms describing the same idea often leads to misunderstanding. This is particularly harmful when talking about scientific ideas. As an example, the "adiabatic demagnetization" was a cryogenic process developed independently, which is equivalent to what we understand today as magnetocaloric effect. Given the variety of ways a material can be mechanically deformed, the vocabulary related to mechanocaloric effects became large. Generically speaking, mechanocaloric effects are often called just "thermal effects" or

"thermoelastic effects". Regarding polymeric materials and composites, "thermo-viscoelastic" properties are subject of study, result of their characteristic viscoelastic behavior. In the field of ceramic materials, "flexocaloric" and "piezocaloric" describe specific measurement conditions, from which thermal responses are monitored. To avoid unnecessary multiplicity in terminology, we propose *i*-caloric effect as a unified label that describes any thermal response induced by external fields, comprising magnetocaloric effect, electrocaloric effect and mechanocaloric effect.

## 2.3. Thermodynamics of *i*-caloric effects

Let us consider the effect of a field $X$ acting on a material, changing the corresponding conjugated specific quantity $Y$. The specific work done on the material by the external field is

$$W = \lambda \int X dY \qquad \text{Eq. (1)}$$

and $\lambda = \pm 1$ indicates the effect of the external field on the internal energy. For instance, pressure increases the internal energy of a gas by reducing its volume (in this case, $\lambda = -1$); on the other hand, an electric field increases the internal energy of a dielectric by increasing its polarization (in this case, $\lambda = +1$). In order to understand the thermal response when an external field is applied on a material, let us regard the intensive internal energy ($U$) of a general closed solid-state system, without change in composition, as follows from the first law of thermodynamics:

$$dU = TdS + \sum_{i=1}^{n} \lambda_i X_i dY_i \qquad \text{Eq. (2)}$$

where $T$ is the absolute temperature, $S$ is the specific entropy. We can view the internal energy as a function of the intensive variables $U = U(S, Y_1, Y_2, \ldots, Y_n)$ and, since $dU$ is a total differential, the variables $X_i$ for the generalized external fields relates to the generalized intensive variables by

$$X_i = \lambda_i \left(\frac{\partial U}{\partial Y_i}\right)_{S, Y \neq Y_i} \qquad \text{Eq. (3)}$$

Now, by applying a series of Legendre transformations, we introduce the Gibbs free energy, which is a function of the intensive variables, $G = G(T, X_1, X_2, \ldots, X_n) \equiv U - TS - \lambda_1 X_1 Y_1 - \lambda_2 X_2 Y_2 - \cdots - \lambda_n X_n Y_n$. Its differential form is

$$dG = -SdT - \sum_{i=1}^{n} \lambda_i Y_i dX_i \qquad \text{Eq. (4)}$$

while the total differential of $G$ is

$$dG = \left(\frac{\partial G}{\partial T}\right)_X dT + \left(\frac{\partial G}{\partial X_1}\right)_{T, X \neq X_1} dX_1 + \cdots + \left(\frac{\partial G}{\partial X_n}\right)_{T, X \neq X_n} dX_n \qquad \text{Eq. (5)}$$

and we have by comparing Eq. (4) to Eq. (5):

$$S = -\left(\frac{\partial G}{\partial T}\right)_X \qquad \text{Eq. (6.1)}$$

$$Y_i = -\lambda_i \left(\frac{\partial G}{\partial X_i}\right)_{T, X \neq X_i} \qquad \text{Eq. (6.2)}$$

Then, if we apply the Euler reciprocity relation in Eq. (6), we are able to construct a generalized Maxwell equation as follows:

$$\left(\frac{\partial S}{\partial X_i}\right)_{T,X\neq X_i} = \lambda_i \left(\frac{\partial Y_i}{\partial T}\right)_X \qquad \text{Eq. (7.1)}$$

$$\lambda_i \left(\frac{\partial Y_j}{\partial X_i}\right)_{T,X\neq X_i} = \lambda_j \left(\frac{\partial Y_i}{\partial X_j}\right)_{T,X\neq X_j} \qquad \text{Eq. (7.2)}$$

If we assume that $S = S(T, X_1, X_2, ..., X_n)$, the total differential of $S$ takes the form

$$dS = \left(\frac{\partial S}{\partial T}\right)_X dT + \left(\frac{\partial S}{\partial X_1}\right)_{T,X\neq X_1} dX_1 + \cdots + \left(\frac{\partial S}{\partial X_n}\right)_{T,X\neq X_n} dX_n \qquad \text{Eq. (8)}$$

and, from the second law of thermodynamics for reversible processes, the term $\left(\frac{\partial S}{\partial T}\right)_X$ from Eq. (8) is intrinsically linked to the specific heat under constant fields ($c_X$) as

$$c_X = T \left(\frac{\partial S}{\partial T}\right)_X \qquad \text{Eq. (9)}$$

Using Eq. (7.1) and Eq (9), Eq. (8) can be rewritten as

$$dS = \frac{c_X}{T} dT + \sum_{i=1}^{n} \lambda_i \left(\frac{\partial Y_i}{\partial T}\right)_{X_i} dX_i \qquad \text{Eq. (10)}$$

Thus, to quantify the thermal response ($\Delta T_S$ and $\Delta S_T$), we have to input in Eq. (10) that: $dS = 0$ for an adiabatic process, and $dT = 0$ for an isothermal process. Then,

$$\Delta T_S = -\sum_{i=1}^{n} \lambda_i \int_{\Delta X_i} \frac{T}{c_X} \left(\frac{\partial Y_i}{\partial T}\right)_{X_i} dX_i , \qquad \text{Eq. (11.1)}$$

$$\Delta S_T = \sum_{i=1}^{n} \lambda_i \int_{\Delta X_i} \left(\frac{\partial Y_i}{\partial T}\right)_{X_i} dX_i , \qquad \text{Eq. (11.2)}$$

where $\Delta X$ is the external field change from the initial external field ($X_{initial}$) to the final external field ($X_{final}$), i.e, $\Delta X \equiv X_{final} - X_{initial}$.

## 3. ARGUMENTS FOR NORMALIZATION

The growing interest on *i*-caloric effects has resulted in a vast literature, with various particularities in terminologies and different ways of presenting data. Mostly, it seems to us that an effort is required for a good comprehension of reported works concerning this area of knowledge. With this idea in mind, we have pointed out some topics in the following subsections, which aim to address some issues of this vast literature and propose ways to deal with them.

### 3.1. Abbreviations for *i*-caloric effects

For *i*-caloric effects, we strongly suggest the abbreviations proposed in Table 1, where "*i*" stands for the external field corresponding to each *i*-caloric effect. The main justification is, recurrently, the abbreviations in several works are confusing (e.g., eCE and ECE for elastocaloric and electrocaloric effects, or mCE and MCE for mechanocaloric and magnetocaloric effects). In a sense, these confusions are eliminated if we keep in mind that the first letter used in the abbreviation is related to the external field that produces the *i*-caloric effect. Also, as marked in the subdivision of the σ-CE, the sub-indices "*e*", "*b*", and "*t*" are for elastocaloric ($\sigma_e$-CE), barocaloric ($\sigma_b$-CE), and torsiocaloric effects ($\sigma_t$-CE), respectively.

## 3.2. Adiabatic temperature change ($\Delta T_S$) and isothermal entropy change ($\Delta S_T$)

We have seen that the *i*-caloric effects are characterized by $\Delta T_S$ (Eq. 10.1) or $\Delta S_T$ (Eq. 10.2), then each material will present $\Delta T_S$ or $\Delta S_T$ values which depend on the process performed. If we want to compare two different materials, in their respective *i*-caloric properties ($\Delta T_S$ or $\Delta S_T$), the ideal situation would be to compare their $\Delta T_S$ or $\Delta S_T$ values in the same applied field and temperature; but these direct comparisons are not usually possible, because the available data do not let us to do that. As a consequence, it is very common to show $\Delta T_S$ or $\Delta S_T$ values normalized by applied fields, but such comparison may lead to erroneous conclusions.

Considering the equivalence of the thermodynamic equations for different *i*-caloric effects, we can understand the ideas stated above taking only $\sigma_b$-CE as example. Fig. 3 displays $\Delta T_S$ of vulcanized natural rubber (VNR) (Bom et al., 2018) and acetoxy silicone rubber (ASR) (Imamura et al., 2017), both measured at two different temperatures (223 and 313 K), in different pressure changes ($\Delta \sigma_b$). The first observation is that the temperature under consideration can influence the comparison. For instance, if we choose to compare both at the same temperature, the $\Delta T_S$ values of ASR are always higher than the $\Delta T_S$ values of VNR, at any pressure change; but if the $\Delta T_S$ values of ASR and VNR were compared in different isotherms, they can lead to different conclusions. For example, if we take the $\Delta T_S$ of ASR at 223 K and compare it to the $\Delta T_S$ of VNR at 313 K, ASR only presented better results for $\Delta \sigma_b < 0.2$ GPa. Concerning normalization of $\Delta T_S$ at the same isotherm (i.e., $\Delta T_S / \Delta \sigma_b$ at $T$), we may have cases where normalization would lead us to conclude VNR presents better results than ASR (e.g. at 223 K, $\Delta T_S / \Delta \sigma_b$ of NVR at $\Delta \sigma_b = 0.87$ GPa is ~59 K GPa$^{-1}$, but for ASR at $\Delta \sigma_b = 0.39$ GPa is ~45 K GPa$^{-1}$).

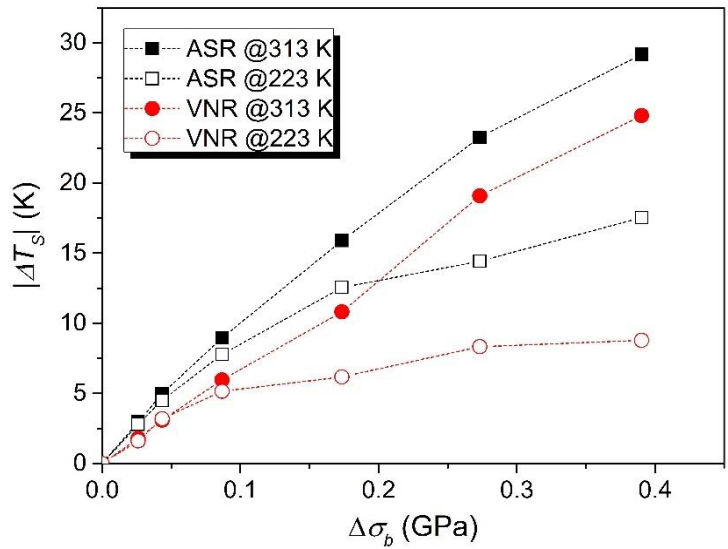

**Figure 3: $\Delta T_S$ vs. $\Delta \sigma_b$ for VNR (Bom et al., 2018) and ASR (Imamura et al., 2017) measured at 223 and 313 K.**

One way to eliminate such possibility of erroneous conclusion – and even to ensure a fairer comparison between different materials under different conditions such $T$ or $\Delta X$ – is to plot curves of $\Delta T_S / \Delta X$ vs. $\Delta T_S$, indicating their respective temperatures. See Fig. 4 as an example of $\Delta T_S / \Delta \sigma_b$ vs. $\Delta T_S$ for selected materials.

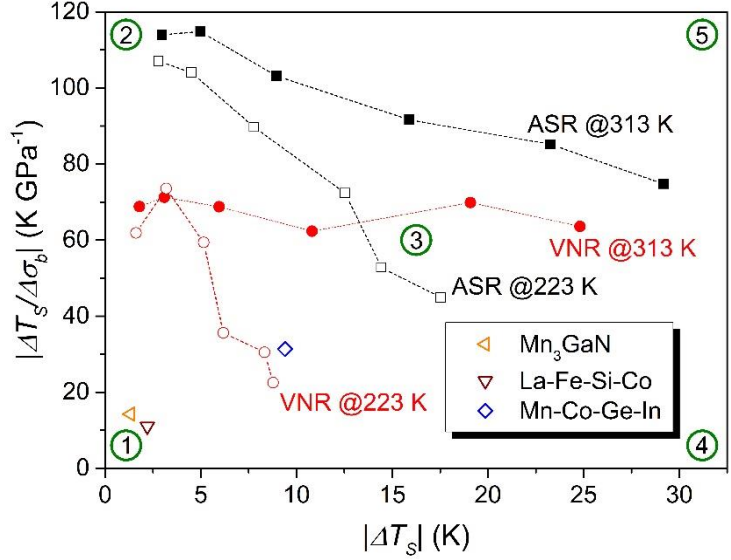

**Figure 4:** Normalization $\Delta T_S/\Delta \sigma_b$ vs. $\Delta T_S$ for some materials. VNR obtained at 223 and 313 K, ASR obtained at 223 and 313 K, $Mn_3GaN$ obtained at 288 K, $Gd_5Si_2Ge_2$ obtained at 272 K, La-Fe-Si-Co obtained at 233 K, and Mn-Co-Ge-In obtained at 298 K (Bom et al., 2018; Imamura et al., 2017; Mañosa et al., 2011; Matsunami et al., 2015; Wu et al., 2015). Circles 1 to 5 indicate regions of the graph, which are discussed in the text.

It is notwithstanding that $\Delta T_S/\Delta \sigma_b$ vs. $\Delta T_S$ plots are more sensitive to compare similar values of $\Delta T_S$. When we observe the normalization of VNR or ASR and compare them to their absolute value of $\Delta T_S$, subtle differences assume distinctive bahavior (especially of lower $\Delta \sigma_b$); this behavior does not appear in Fig. 3. Moreover, this type of plot organizes different materials in their respective similarities for *i*-caloric applications (see circles 1 to 5 in Fig. 4):

- Circle 1: region near the origin (low $\Delta T_S/\Delta \sigma_b$ values with low $\Delta T_S$ values). Not desirable; $Mn_3GaN$ and La-Fe-Si-Co as examples.

- Circle 2: region near the upper-left corner (high $\Delta T_S/\Delta \sigma_b$ values with low $\Delta T_S$ values). Some materials can exhibit striking normalized quantities under low applied fields, which do not guarantee high $\Delta T_S$. [TPrA][Mn(dca)$_3$] (Bermúdez-García et al., 2017), for example, reaches ~600 K GPa$^{-1}$, but its $\Delta T_S$ value is 4.1 K.

- Circle 3: central region (moderate $\Delta T_S/\Delta \sigma_b$ values with moderate $\Delta T_S$ values).

- Circle 4: region near the down-right corner ($\Delta T_S/\Delta \sigma_b$ values with high $\Delta T_S$ values). This region is the opposite to what was described for region 2. Some materials can exhibit striking $\Delta T_S$ values, but the applied fields are so high (thus reducing the normalizations), which is almost inapplicable.

- Circle 5: region near the upper-right corner (high $\Delta T_S/\Delta \sigma_b$ values with high $\Delta T_S$ values). This is the ideal region of interest, because $\Delta T_S/\Delta \sigma_b$ and $\Delta T_S$ must have high values, otherwise only $\Delta T_S/\Delta \sigma_b$ vs. $T$ or $\Delta T_S$ vs. $\Delta \sigma_b$ data can mask the results by themselves.

The same idea of normalization can be extended for $\Delta S_T$ data, plotting $\Delta S_T/\Delta X$ vs. $\Delta S_T$ curves at different temperatures.

### 3.3. Figures of merit

There are different figures of merit that try to predict the most promising *i*-caloric material for a specific device. Recently, Griffith et al. (2018) discussed several figures of merit for *h*-CE, and proposed the *Temperature average Entropy Change* (*TEC*) as

$$TEC(\Delta T_{h-c}) = \frac{1}{\Delta T_{h-c}} \max\left\{\int_{T_{mid}-\frac{\Delta T_{h-c}}{2}}^{T_{mid}+\frac{\Delta T_{h-c}}{2}} |\Delta S_T|\, dT\right\} \quad \text{Eq. (12)}$$

where $\Delta T_{h-c}$ is the difference between the hot reservoir and the cold reservoir ($\Delta T_{h-c} \equiv T_{hot} - T_{cold}$) of the device, and $T_{mid}$ is the temperature which maximizes the integral. This method is very similar to another which is based on the maximization of *Refrigerant Capacity* (*RC*), and had been previously proposed and applied for magnetocaloric materials (Carvalho et al., 2013). Since *RC* is defined as

$$RC = \int_{T_{cold}}^{T_{hot}} |\Delta S_T|\, dT \quad \text{Eq. (13)}$$

we can conclude by comparing Eq. (12) and Eq. (13) that the only difference between these two methods is that *TEC* is divided by $\Delta T_{h-c}$. In other words, there is always a given $T_{mid}$ for which the maximization of *RC* is $RC_{max} = TEC \times \Delta T_{h-c}$.

Another figure of merit is the *Normalized Refrigerant Capacity* (*NRC*) (Bom et al., 2018; Imamura et al., 2017), which is nothing more than the *RC* divided by the external field change:

$$NRC = \frac{1}{\Delta X} \int_{T_{cold}}^{T_{hot}} |\Delta S_T|\, dT \quad \text{Eq. (14)}$$

Although Eq. (12) to Eq. (14) is extended to different *i*-caloric materials, none of them is sufficient to ensure the best choice by themselves. Due to the different variables in each methods of presenting data, only a single figure is not the most appropriate, then we suggest a set of figures, as we display in Fig. 5a-c. Going further, if necessary, maybe 3D graphics could be even more informative.

Lastly, the *Coefficient of Performance* (*COP*) is another figure of merit widely used for *i*-caloric materials. This parameter can be defined as

$$COP = \left|\frac{Q}{W}\right| \quad \text{Eq. (15)}$$

where $Q = T\Delta S_T$ is the heat that can be removed from the material in a certain temperature, and $W$ is the specific work required for that. Since *COP* is dimensionless, it can be used to compare different *i*-caloric effects between them. Attention should be drawn during the calculation of the specific work (see Table 2).

**Table 2: Specific work equations used during the calculation of COP, for each type of i-caloric effect.**

| *i*-caloric effect | Specific work* ($W$) |
|---|---|
| Magnetocaloric (*h*-CE) | $\int_{M_i}^{M_f} H\, dM$ |
| Electrocaloric (*e*-CE) | $\int_{P_i}^{P_f} E\, dP$ |
| Mechanocaloric ($\sigma$-CE) | $\frac{1}{\rho_0}\int_{\varepsilon_i}^{\varepsilon_f} \sigma\, d\varepsilon$ |

* The subscripts *i* and *f* on the limits of the integral indicate initial and final states, respectively.

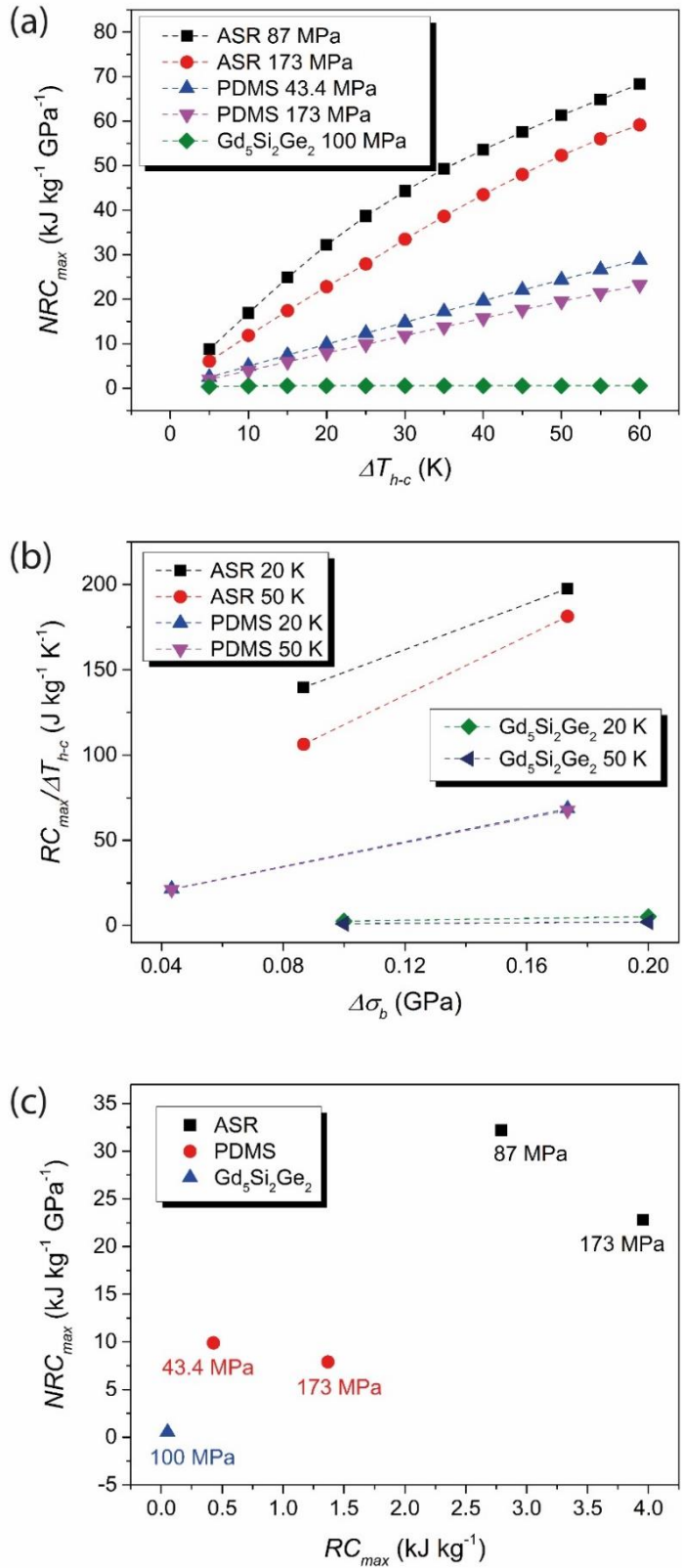

**Figure 5: Performance coefficients for ASR, PDMS and Gd$_5$Si$_2$Ge$_2$.** (a) $NRC_{max}$ vs. $\Delta T_{h-c}$ for ASR ($\Delta\sigma_b$ = 87 and 173 MPa) (Imamura et al., 2017), PDMS ($\Delta\sigma_b$ = 43.4 and 173 MPa) (Carvalho et al., 2018), and Gd$_5$Si$_2$Ge$_2$ ($\Delta\sigma_b$ = 100 MPa) (Yuce et al., 2012); (b) $RC_{max}/\Delta T_{h-c}$ vs. $\Delta\sigma$ with $\Delta T_{h-c}$ = 20 and 50 K for ASR ($\Delta\sigma_b$ = 87 and 173 MPa) (Imamura et al., 2017), PDMS ($\Delta\sigma_b$ = 43.4 and 173 MPa) (Carvalho et al., 2018), and Gd$_5$Si$_2$Ge$_2$ ($\Delta\sigma_b$ = 100 and 200 MPa) (Yuce et al., 2012). (c) N$RC_{max}$ vs. $RC_{max}$ for ASR ($\Delta\sigma_b$ = 87 and 173 MPa) (Imamura et al., 2017), PDMS ($\Delta\sigma_b$ = 43.4 and 173 MPa) (Carvalho et al., 2018), and Gd$_5$Si$_2$Ge$_2$ ($\Delta\sigma_b$ = 100 MPa) (Yuce et al., 2012).

## 4. FINAL REMARKS

The vast knowledge accumulated on *i*-caloric effects over the past decades demonstrate the relevance of this field for the current research towards novel cooling technologies. However, this intense activity has also brought up issues regarding the lack of standards for researchers to evaluate, compare and report their results. The present effort is the first attempt to unify the knowledge built by the *i*-caloric community, covering the topics: i) a general thermodynamic formalism for *i*-caloric effects; ii) a set of figures to display and compare the results; iii) new terminologies and abbreviations. Although several topics were addressed, there is still a lot work to be done, and we hope the present study encourage the community to take part in the discussion.

## ACKNOWLEDGEMENTS

The authors acknowledge financial support from FAPESP (project number 2016/22934-3), CNPq, CAPES, LNLS and CNPEM.

## NOMENCLATURE

| | | | |
|---|---|---|---|
| $\Delta T_S$ | adiabatic temperature change | $X$ | external field |
| $\Delta S_T$ | isothermal entropy change | $Y$ | thermodynamic conjugate of $X$ |
| $H$ | magnetic field | $W$ | work |
| $E$ | electric field | $U$ | internal energy |
| $\sigma$ | Cauchy stress tensor | $T$ | temperature |
| $M$ | magnetization | $S$ | entropy |
| $P$ | polarization | $G$ | Gibbs free energy |
| $\varepsilon$ | strain | $c_X$ | specific heat under constant fields |
| $h$-CE | magnetocaloric effect | VNR | vulcanized natural rubber |
| $e$-CE | electrocaloric effect | ASR | acetoxy silicone rubber |
| $\sigma$-CE | mechanocaloric effect | $\Delta \sigma_b$ | pressure change |
| $\sigma_e$-CE | elastocaloric effect | TEC | temperature average entropy change |
| $\sigma_b$-CE | barocaloric effect | RC | refrigerant capacity |
| $\sigma_t$-CE | torsiocaloric effect | $\Delta T_{h-c}$ | temperature difference between hot and cold reservoirs |
| $\sigma_{ii}$ | components of stress tensor | NRC | normalized refrigerant capacity |
| $p$ | pressure | COP | coefficient of performance |